\documentclass{epl}

\newcommand{\bsigma}{\mbox{\boldmath$\sigma$}}
\newcommand{\ep}{\varepsilon}
\newcommand{\nn}{\nonumber\\}

\title{Can the spin-orbit interaction break the channel degeneracy of
the two-channel orbital Kondo problem?}
\shorttitle{Can the spin-orbit interaction break...}

\author{O. \'Ujs\'aghy\inst{1} \and A. Zawadowski\inst{1,2}}
\institute{
  \inst{1} Institute of Physics and Research Group of Hungarian
Academy of Sciences, Budapest University of Technology and Economics - 
H-1521 Budapest, Hungary\\
  \inst{2} Research Institute for Solid State Physics - POB 49, H-1525
Budapest, Hungary}

\begin{document}

\maketitle

\begin{abstract}
  Two-level systems (TLS) interacting with conduction electrons are possibly 
  described by the two-channel Kondo Hamiltonian. In this case the channel
  degeneracy is due to the real spin of the electrons. The possibility
  of breaking that degeneracy has interest on his own. In fact, we
  show that the interaction of the conduction electrons with a
  spin-orbit scatterer nearby the TLS leads to the breaking
  of the channel degeneracy only in the case of electron-hole symmetry 
  breaking. The generated channel symmetry breaking TLS-electron couplings 
  are, however, too weak to result in any observable effects.
  Our analysis is also relevant for heavy fermion systems.
\end{abstract}

In the general form of the orbital Kondo model 
 a single particle is moving between two localized orbitals and  is
interacting with the conduction electrons in a metal. 
This orbital Kondo model has been justified by a detailed  scaling
analysis~\cite{Anderson,Cox,VZ,ZarVlad}, 
though it is presently unclear if the two-channel Kondo fixed point 
can be experimentally reached in the case of TLS's~\cite{Aleiner,Kagan,BZ}.
The particle can be  an atom or a group of atoms. Similar orbital models
emerge in the context of 4f heavy fermion impurities~\cite{Cox}.
In these models the real electronic spin variable does not occur in the 
coupling constants, thus there is a spin degeneracy in the variables of the 
particles and all interaction  terms are diagonal in the conduction electron 
spin. 
In realistic materials, however, spin-orbit interaction is always present, 
and it always induces cross-scattering between different spin orientations. 
It is, therefore, a fundamental question, whether spin-orbit interaction 
can break this channel symmetry and invalidate the 2CK description or not.
 
In this paper we examine the possibility of breaking the channel
degeneracy of the orbital Kondo problem due to the interaction of the
conduction electrons with a spin-orbit scatterer nearby the TLS, using
the renormalization group method in leading logarithmic order. It
turns out, that in case of electron-hole symmetry the
spin-orbit interaction has no effect on the two-channel Kondo
behavior. In contrary, in case of electron-hole symmetry breaking,
new, relevant channel symmetry breaking couplings are generated
between the TLS and the conduction electrons~\cite{UZVZ}, 
which are driven by the rather small ratio of
the TLS level splitting and the electronic bandwidth. 
As a consequence, despite of its relevance in the RG sense,
this term cannot influence the two-channel behavior
in an observable range of temperature, since the scaling is stopped by the
infrared cutoff (TLS level splitting) long before the corresponding 
crossover is reached.

\begin{figure}
  \onefigure[scale=0.6]{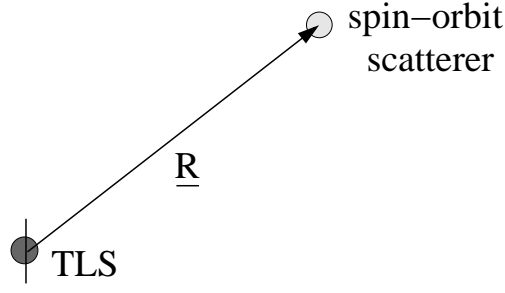}
  \caption{The TLS and the spin-orbit scatterer in a distance {\bf R}.}
\label{fig1}
\end{figure}

We consider a TLS interacting with conduction electrons which are also
interacting with a spin-orbit scatterer at a position {\bf R} with
respect to the TLS (see fig.~\ref{fig1}).  
The TLS-conduction
electron system is described by the usual Hamilton operator~\cite{Zawa}
\begin{equation}
\label{Ham}
  H_{TLS-el}=\sum\limits_{k,l,m,\sigma}\!\!\varepsilon_k \,
  a_{klm\sigma}^\dagger a_{klm\sigma}+\Delta_0\sigma^x_{TLS}+
  \Delta\sigma^z_{TLS}+
  \sum\limits_{\scriptstyle i=x,y,z
    \atop{l, l'\atop{k, k',\sigma}}} V^i\sigma^i_{TLS}
  a^\dagger_{kl\sigma}(\sigma^i_{el})_{ll'} a_{k'l'\sigma}
 \end{equation}
where $a_{klm\sigma}^\dagger$ creates an electron with momentum $k$,
angular momentum $l,m$ and spin $\sigma$, $\sigma^i$ stand for the
Pauli matrices, $\Delta_0$ and $\Delta$ are the spontaneous transition
and the energy splitting between the two TLS states, respectively.
Choosing the z axis in an appropriate way and assuming axial symmetry, 
the TLS is strongly coupled only
to a reduced number of channels e.g. to those with azimuthal quantum number 
$m=0$ of the conduction electrons, thus the
$m$ indices are dropped and only two angular momenta $l=0,1$ are 
kept~\cite{Zawa}.

\begin{figure}
  \onefigure[width=13cm]{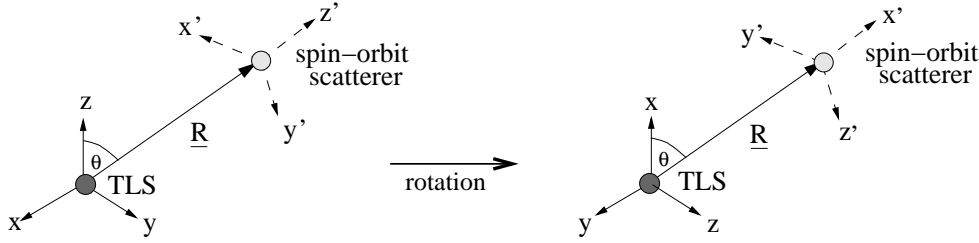}
  \caption{The TLS ($\mathrm{x,y,z}$) and the local
    ($\mathrm{x',y',z'}$) frame.
    The simultaneous rotations to the new TLS and the new local frame
    are also illustrated.}
\label{fig2}
\end{figure}

We describe the spin-orbit scatterer by an Anderson-like ($l=2$) 
model~\cite{UZaniz} with parameters $\varepsilon_0$ and $V_{kml'm'}({\bf R})$
as~\cite{UZaniz} 
\begin{equation}
H_{s-o}=\varepsilon_0\sum\limits_{m\sigma}\!\!b^{\dagger}_{m\sigma}
b_{m\sigma} +\sum\limits_{kl'mm^\prime \sigma}\!\!\biggl(
V_{kml'm'}({\bf R})\,b^{\dagger}_{m\sigma} a_{kl'm'\sigma} + {\rm h.c.}
\biggr)
+\lambda\sum\limits_{\scriptstyle mm' \atop \scriptstyle
\sigma\sigma'}\!\!\langle m|
{\bf L}|m'\rangle \langle\sigma|\bsigma|\sigma'\rangle
b^{\dagger}_{m\sigma} b_{m'\sigma'}
\end{equation}
where $b^\dagger_{m\sigma}$ creates an electron on the spin-orbit
scatterer orbital labeled by the quantum numbers $m,\sigma$ and
$\lambda$ is the strength of the spin-orbit interaction. The
hybridization matrix element, $V_{kml'm'}({\bf R})$ depends on the relative
position of the two coordinate systems with origin at the TLS and the
spin-orbit scatterer, respectively~\cite{UZaniz} (see fig.~\ref{fig2}).

The calculation of the correction to the electron Green's function due
to spin-orbit interaction was performed in a similar way as in 
Section III of ref.~\cite{UZaniz}. 
The first order correction $\delta G^{(1)}$ in the spin-orbit coupling is 
given by 
\begin{equation}
\label{korr}
\delta G^{(1)}_{l l'\atop\sigma\sigma'} (0,0,i\omega_n)=G^{(0)}
(0,{\bf R},i\omega_n) g
\sigma^y_{l'l}\sigma^y_{\sigma'\sigma} G^{(0)} ({\bf R},0,i\omega_n)
\end{equation}
where g depends on the parameters of the spin-orbit scatterer's
d-level, the strength of the spin-orbit interaction, the angle
$\theta$ between the TLS axis and {\bf R}, and on the distance $R$
between the TLS and the spin-orbit scatterer, like
$\sim \frac{1}{(k_F R)^3}$ in leading order. 
In eq.~(\ref{korr}) the orbital momentum
variables of the electron are in the "TLS frame" (x,y,z)
($\mathrm{z}\parallel$TLS axis), whereas the spin variables are in a
"local frame" (x',y',z') ($\mathrm{z}'\parallel {\bf R}$) (see
fig.~\ref{fig2}).
 
After simultaneous rotation of both coordinate systems to a new TLS
($\mathrm{x}\parallel$TLS axis) and a new local ($\mathrm{x}'\parallel {\bf
  R}$) frame (see fig.~\ref{fig2}) the scattering amplitude contained
by $\delta G^{(1)}$ can be summed up
to infinite order (i.e. infinite number of scatterings on the same
spin-orbit scatterer is considered), resulting in
\begin{equation}
\delta G_{l l'\atop\sigma\sigma'} (0,0,i\omega_n)=
G^{(0)} (0,{\bf R},i\omega_n) \frac{g\sigma^z_{l'l}
\sigma^z_{\sigma'\sigma}}{1-g
  l\sigma G^{(0)} ({\bf R},{\bf R},i\omega_n)} G^{(0)} ({\bf R},0,i\omega_n).
\end{equation}

Calculating the corresponding change in the conduction electron
density of states in first order in $g$ and using the linearized
dispersion $k=k_F+\frac{\ep}{v_F}$ near the Fermi level, we get for
the spin-dependent part
\begin{equation}
\frac{\delta\rho_R
  (\omega\approx 0)}{\rho_0}=\Lambda\sigma^z_{l'l}\sigma^z_{\sigma'\sigma}
\end{equation}
where $\Lambda$ depends on $g$, the conduction
electron density of states at Fermi level for one spin direction $\rho_0$,  
and in leading order it is 
$\sim \frac{1}{(k_F R)^5}$, therefore only the first neighboring atoms 
around the TLS give non-negligible contribution. 

We used the above result to examine the TLS-conduction electron system in
case of finite $\Lambda$.
The new TLS-electron couplings, obtained by introducing the dimensionless 
TLS-electron couplings and taking into account the above changes in the 
conduction electron density of states, are
\begin{eqnarray}
v^x&=&\rho_0 V^x \longrightarrow {\tilde
v^x}=v^x\sqrt{1+\Lambda}\sqrt{1-\Lambda}\nonumber \\
v^y&=&\rho_0 V^y \longrightarrow {\tilde
  v^y}=v^y\sqrt{1+\Lambda}\sqrt{1-\Lambda}
\end{eqnarray}
where the different signs in front of $\Lambda$'s are due to the off-diagonal
behavior in $l$ and $l'$.
Then the term with coupling $\sim v^z$ 
in the Hamiltonian (\ref{Ham}) is replaced by the spin dependent term
\begin{equation}
v^z\sigma^z_{ll'}\sigma^z_{\alpha\alpha'}\delta_{\sigma\sigma'}\longrightarrow
  v^z \sigma^z_{\alpha\alpha'} (\sigma^z_{ll'}\delta_{\sigma\sigma'} +
  \Lambda \delta_{ll'} \sigma^z_{\sigma\sigma'}),
\end{equation}
where $v_z=\rho_0 V_z$, and $l,l'$, $\sigma,\sigma'$ correspond to the
orbital momentum and the real spin of the conduction electrons,
respectively, and $\alpha,\alpha'$ label the TLS states.

To investigate the possibility of breaking the channel degeneracy by
the spin-orbit interaction, we performed a scaling analysis in leading
logarithmic approximation for general,
$v^\mu_{\nu\rho}\sigma^\mu_{\alpha'\alpha}
\sigma^\nu_{l'l}\sigma^\rho_{\sigma'\sigma}$ couplings where
$\mu,\nu,\rho=0,x,y,z$, and $\sigma^0$ is the unity matrix.
In the calculation we used
$\rho(\ep)=\rho_0 (1+\frac{\alpha\ep}{D_0})$ for the conduction
electron density of states in order to account for 
the electron-hole symmetry breaking in a simple way~\cite{UZVZ,Nozi}, 
(where $D_0$ is in the range of the electronic
bandwidth which is not subject of scaling and $|\alpha|<1$).
\begin{figure}
  \onefigure[scale=0.5]{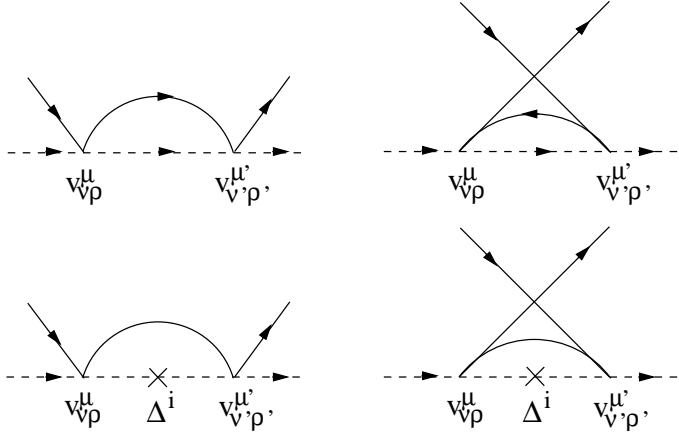}
  \caption{The diagrams generating the leading logarithmic scaling
    equations. The solid and dotted lines represent the conduction
    electrons and the TLS, respectively, and the crosses indicate the
    TLS level splitting.}
\label{fig3}
\end{figure}

The generating diagrams of the leading logarithmic scaling equations
are shown in fig.~\ref{fig3} and the corresponding scaling equations
read as
\begin{eqnarray}
\frac{\partial v^\mu_{\nu\rho}}{\partial
  x}=&-&\sum\limits_{{\mu_1,\mu_2=0,x,y,z\atop\nu_1,\nu_2=0,x,y,z}\atop
\rho_1,\rho_2=0,x,y,z}
  \biggl \{i v^{\mu_1}_{\nu_1\rho_1} v^{\mu_2}_{\nu_2\rho_2} 
\varepsilon^{\mu_2\mu_1\mu}\bigl (\varepsilon^{\nu_1\nu_2\nu} 
\varepsilon^{\rho_1\rho_2\rho}-\varepsilon^{\nu_2\nu_1\nu} 
\varepsilon^{\rho_2\rho_1\rho}\bigr )\nonumber \\
&+&\sum\limits_{i=x,y,z\atop \mu'=0,x,y,z}\Delta^i\frac{\alpha}{D} 
v^{\mu_1}_{\nu_1\rho_1} v^{\mu_2}_{\nu_2\rho_2}
\varepsilon^{\mu_2 i\mu'} \varepsilon^{\mu'\mu_1\mu}\bigl (
\varepsilon^{\nu_1\nu_2\nu}\varepsilon^{\rho_1\rho_2\rho}
+\varepsilon^{\nu_2\nu_1\nu} 
\varepsilon^{\rho_2\rho_1\rho}\bigr )\biggl\}
\end{eqnarray}
where $\varepsilon^{\mu_1\mu_2\mu_3}$ is the usual Levi-Civita symbol
for $\mu_1,\mu_2,\mu_3=x,y,z$, $\varepsilon^{0\mu_1\mu_2}=
\varepsilon^{\mu_1 0\mu_2}=\varepsilon^{\mu_1\mu_2
  0}=-i\delta_{\mu_1\mu_2}$, and $x=\ln\frac{D_0}{D}$.  
We can see immediately that in the presence of 
electron-hole symmetry (i.e. $\alpha=0$) we reproduce the
usual TLS-electron scaling equations, thus the spin-orbit interaction
cannot influence the behavior of the TLS-electron system in this case. 

Together with the initial conditions ($v^s_{p0}(0)=\delta_{sp}{\tilde
  v_s}$ for $s,p=x,y,z$, $v^z_{0z}(0)=\Lambda v^z$ and the other $v$'s
are zero), the above scaling equation system is closed for the
subspace $\rho=0,z$, thus we can restrict the general equations to
those values and then we divide the relevant couplings to spin
independent and spin dependent parts as
\begin{eqnarray}
v^{\mu}_{\nu}&:=&\frac{v^\mu_{\nu\uparrow}+v^\mu_{\nu\downarrow}}{2}
=v^\mu_{\nu 0}\nn
\delta v^{\mu}_{\nu}&:=&\frac{v^\mu_{\nu\uparrow}-v^\mu_{\nu\downarrow}}{2}
=v^\mu_{\nu z}
\end{eqnarray}
where $v^\mu_{\nu\uparrow}$ and $v^\mu_{\nu\downarrow}$ are the
couplings for up and down electron spins, respectively. The scaling equations
for the spin independent and spin dependent couplings then read for $s,p=x,y,z$
\begin{eqnarray}
  \frac{\partial v^0_0}{\partial x}&=& - 4 \sum\limits_{i=x,y,z}
  \Delta^i\frac{\alpha}{D_0} (v^0_0 v^i_0+v^0_z v^i_z) \nonumber \\
  \frac{\partial v^s_0}{\partial
  x}&=&-4\sum\limits_{i=x,y,z}\Delta^i\frac{\alpha}{D_0} (v^s_0 v^i_0+v^s_z
  v^i_z)
  + 2\Delta^s\frac{\alpha}{D_0} (\sum\limits_{q=x,y,z} (v^q_0 v^q_0+v^q_z
  v^q_z)-v^0_0 v^0_0-v^0_z v^0_z)\nonumber \\
  \frac{\partial v^0_p}{\partial
    x}&=&-4\sum\limits_{i=x,y,z}\Delta^i\frac{\alpha}{D_0} (v^0_0 v^i_p+v^0_p
  v^i_0) \nonumber \\
  \frac{\partial v^s_p}{\partial
    x}&=&2 \sum\limits_{i_1,i_2=x,y,z\atop j_1,j_2=x,y,z} v^{i_1}_{j_1} 
  v^{i_2}_{j_2}
  \ep^{i_1 i_2 s} \ep^{j_1 j_2 p}
  -4\sum\limits_{i=x,y,z}\Delta^i\frac{\alpha}{D_0} (v^s_0 v^i_p+v^s_p
  v^i_0)\nn
  &+&4\Delta^s\frac{\alpha}{D_0} (\sum\limits_{q=x,y,z} v^q_0 v^q_p-v^0_0
  v^0_p)
\end{eqnarray}
and
\begin{eqnarray}
  \frac{\partial (\delta v^0_0)}{\partial
    x}&=&-4\sum\limits_{i=x,y,z}\Delta^i\frac{\alpha}{D_0} 
  (v^0_0 \delta v^i_0+v^0_z
  \delta v^i_z+\delta v^0_0 v^i_0+\delta v^0_z v^i_z) \nonumber \\
  \frac{\partial (\delta v^s_0)}{\partial
    x}&=&-4\sum\limits_{i=x,y,z}\Delta^i\frac{\alpha}{D_0} 
  (v^s_0 \delta v^i_0+v^s_z
  \delta v^i_z+\delta v^s_0 v^i_0+\delta v^s_z v^i_z)\nn
  &+&4 \Delta^s\frac{\alpha}{D_0} (\sum\limits_{q=x,y,z} (v^q_0 \delta
  v^q_0+v^q_z\delta v^q_z)-v^0_0\delta v^0_0-v^0_z\delta
  v^0_z)\nonumber \\
  \frac{\partial (\delta v^0_p)}{\partial
    x}&=&-4\sum\limits_{i=x,y,z}\Delta^i\frac{\alpha}{D_0} 
  (v^0_0 \delta v^i_p+v^0_p
  \delta v^i_0+\delta v^0_0 v^i_p+\delta v^0_p v^i_0) \nonumber \\
  \frac{\partial (\delta v^s_p)}{\partial
    x}&=&4\sum\limits_{i_1,i_2=x,y,z\atop j_1,j_2=x,y,z} 
  v^{i_1}_{j_1}\delta v^{i_2}_{j_2}
  \ep^{i_1 i_2 s} \ep^{j_1 j_2 p}
  - 4\sum\limits_{i=x,y,z}\Delta^i\frac{\alpha}{D_0} (v^s_0 \delta v^i_p+v^s_p
  \delta v^i_0+\delta v^s_0 v^i_p+\delta v^s_p v^i_0)\nonumber \\
  &+&4\Delta^s\frac{\alpha}{D_0} (\sum\limits_{q=x,y,z} (v^q_0 \delta
  v^q_p+v^q_p\delta v^q_0)-v^0_0\delta v^0_p-v^0_p\delta
  v^0_0)
\end{eqnarray}
where the initial values are 
$v^s_p(0)=\delta_{sp} {\tilde v}_s$,
$v^s_0(0)=0$, $v^0_p(0)=0$, $v^0_0(0)=0$, $\delta v^z_0(0)=\Lambda
v^z$, and the other spin dependent couplings are zero.

After linearization in the spin dependent couplings, the scaling equations
for the spin independent couplings decouple from the others. In leading
order in $\frac{\alpha\Delta}{D_0}$, $\frac{\alpha\Delta_0}{D_0}$ 
($\Delta^x=\Delta$, $\Delta^y=\Delta_0$, $\Delta_z=0$ according to the 
coordinate system used), the equations and, thus, the solutions for the
spin independent couplings are the usual ones~\cite{Cox}
\begin{eqnarray}
  v^0_0 (x)&=&v^0_x (x)=v^0_y (x)=v^0_z (x)=v^z_0 (x)=0\nonumber \\
  v^s_p (x)&=&\delta_{sp} v^p (x)\hskip0.2cm {\rm for}\hskip0.2cm
  s,p=x,y,z,
\end{eqnarray}
except that couplings $v^x_0\sim\frac{\alpha\Delta}{D_0}$,
$v^y_0\sim\frac{\alpha\Delta_0}{D_0}$ are generated.

Assuming that these solutions are
isotropic~\cite{Cox,VZ,ZarVlad} ($v^x (x)=v^y (x)=v^z (x)=\Psi (x)$) as is the 
case around $x=\ln\frac{D_0}{T_K}$, the
equations for the spin dependent couplings in leading order in
$\frac{\alpha\Delta}{D_0}$, $\frac{\alpha\Delta_0}{D_0}$ form a
differential equation system with constant coefficients which can be
solved by first order perturbation theory.  Although the solutions
for most of the spin dependent couplings remain zero
($\delta v^0_\nu$, $\delta v^x_0$, $\delta v^y_0$,
$\delta v^x_x$, $\delta v^x_y$, $\delta v^y_x$, 
$\delta v^y_y$, $\delta v^z_z$)
or unrenormalized ($\delta v^z_0$),
new types of couplings $\delta v^x_z$, $\delta v^y_z$, $\delta
v^z_x$, $\delta v^z_y$ are also generated which are
spin-dependent and relevant (growing like $\Psi
(x)\frac{\alpha\Delta}{D_0}$, $\Psi (x)\frac{\alpha\Delta_0}{D_0}$),
thus in principle they break the channel degeneracy of the two-channel
orbital Kondo problem. It is important to emphasize again, that these
new couplings are generated only if the electron-hole symmetry is
broken.  However, using $\frac{\alpha\Delta}{D_0}$,
$\frac{\alpha\Delta_0}{D_0}\approx 10^{-5}$ in the scaling
equations, they are too small to 
influence the two-channel behavior in an observable range of temperature.


To summarize, in this paper we examined the possibility of channel
degeneracy breaking of the two-channel orbital Kondo problem by the
spin-orbit interaction of the conduction electrons. The calculation
was performed in the TLS model, but our analysis is also relevant for heavy 
fermion systems. It turned out that
in case of electron-hole symmetry breaking, the interaction of the
conduction electrons with a spin-orbit scatterer in a position ${\bf
  R}$ according to the TLS, new, relevant, real spin dependent (thus
channel degeneracy breaking) couplings between TLS and conduction
electrons are generated. However, the corresponding crossover between
the 2CK and 1CK behavior cannot be reached as the factor
$\frac{\alpha\Delta}{D_0}$ or $\frac{\alpha\Delta_0}{D_0}$ is
contained in the scaling equations of the channel degeneracy breaking
terms, which is very small. Thus, the channel symmetry breaking is
driven by $\Delta$ or $\Delta_0$, but the same quantities stop the
scaling long before the crossover is reached. This situation is very
similar to the commutative TLS model with impurity
potential~\cite{Moustakas} 
where the commutative marginal line becomes unstable
due to $\frac{\alpha\Delta_0}{D_0}$, but the scaling region is
restricted also by the infrared cutoff $\Delta_0$~\cite{Nozi}. Thus,
we can conclude, that although the spin-orbit interaction, in
principle, can break the channel degeneracy of the two-channel orbital
Kondo problem, that cannot be relevant in physical systems.

\acknowledgments
We are grateful to G. Zar\'and for useful discussions. 
This work was supported by the OTKA Postdoctoral Fellowship D32819
(O.\'U.), by Hungarian grants OTKA T034243, T038162 and grant
No. RTN2-2001-00440. We acknowledge the financial support of EC Human 
Capital Programme.

\end{document}